\documentclass[useAMS,usenatbib]{mn2e}
\usepackage{mn2e-breakabs}
\usepackage{graphicx}
\usepackage{times}
\newcommand{\Hunit}{\,{\rm km}\,{\rm s}^{-1}\,{\rm Mpc}^{-1}}

\def\la{\mathrel{\mathpalette\fun <}}

\def\fun#1#2{\lower3.6pt\vbox{\baselineskip0pt\lineskip.9pt
        \ialign{$\mathsurround=0pt#1\hfill##\hfil$\crcr#2\crcr\sim\crcr}}}

\def\bfk{\mbox{\bf k}}

\def\bfs{\mbox{\bf s}}

\def\rmd{\mbox{d}}

\def\rd{{\rm d}}

\newcommand{\be}{\begin{equation}}
\newcommand{\ee}{\end{equation}}
\newcommand{\ba}{\begin{eqnarray}}
\newcommand{\ea}{\end{eqnarray}}
\newcommand{\simgt}{\,\hbox{\lower0.6ex\hbox{$\sim$}\llap{\raise0.6ex\hbox{$>$}}}\,}
\newcommand{\simlt}{\,\hbox{\lower0.6ex\hbox{$\sim$}\llap{\raise0.6ex\hbox{$<$}}}\,}

\begin{document}

\title[Modeling small-scale galaxy clustering]{Modeling galaxy clustering on small scales to tighten constraints on dark energy and modified gravity}

\author[Yun Wang]{
  \parbox{\textwidth}{
    Yun Wang\thanks{E-mail: wang@ipac.caltech.edu}}
  \vspace*{4pt} \\
IPAC, California Institute of Technology, Mail Code 314-6, 1200 East California Boulevard, 
Pasadena, CA 91125\\  
  Homer L. Dodge Department of Physics \& Astronomy, Univ. of Oklahoma,
                 440 W Brooks St., Norman, OK 73019, U.S.A.\\
                 }

\date{\today}

\maketitle

\begin{abstract}

We present a new approach to measuring cosmic expansion history and growth rate of large scale structure using
the anisotropic two dimensional galaxy correlation function (2DCF) measured from data; it makes use
of the empirical modeling of small-scale galaxy clustering derived from numerical simulations by Zheng et al. (2013).
We validate this method using mock catalogues, before applying it to the analysis of the CMASS sample from the 
Sloan Digital Sky Survey Data Release 10 (DR10) of the Baryon Oscillation Spectroscopic Survey (BOSS).  
We find that this method enables accurate and precise measurements of cosmic expansion history and growth rate of large scale structure.
Modeling the 2DCF fully including nonlinear effects and redshift space distortions (RSD) in the scale range of 16 to 144 $h^{-1}$Mpc,
we find $H(0.57)r_s(z_d)/c=0.0459 \pm 0.0006$, $D_A(0.57)/r_s(z_d)=9.011 \pm  0.073 $,
and $f_g(0.57)\sigma_8(0.57)=0.476 \pm  0.050$, which correspond to precisions of 1.3\%, 0.8\%, and 10.5\% respectively.
We have defined $r_s(z_d)$ to be the sound horizon at the drag epoch computed using a simple integral,
$f_g(z)$ as the growth rate at redshift $z$, and $\sigma_8(z)$ as the matter power spectrum normalization on $8\,h^{-1}$Mpc scale at $z$.
We find that neglecting the small-scale information significantly weakens the constraints on $H(z)$ and $D_A(z)$, and leads to a biased 
estimate of $f_g(z)$. Our results indicate that we can significantly tighten constraints on dark energy and modified gravity by reliably
modeling small-scale galaxy clustering.

\end{abstract}


\section{Introduction}  \label{sec:intro}

Almost two decades after the first detections of cosmic acceleration  \citep{Riess98,Perl99}, we are still in the dark
about its nature. We don't even know if this cosmic acceleration is caused by dark energy (an unknown energy component in the
Universe), or modified gravity (a modification of general relativity).\footnote{For reviews, see, e.g., 
\cite{Ratra08,Frieman08,Caldwell09,Uzan10,Wang10,Li11,Weinberg12}.}

The distribution of galaxies in the Universe traces cosmic large scale structure, and is a powerful probe of the
nature of cosmic acceleration.
Galaxy clustering enables the measurement of cosmic expansion history in two complementary ways \citep{Blake03,Seo03}: through the
direct measurement of $H(z)$, the Hubble parameter (the cosmic expansion rate ${\rm d\,ln}a(t)/dt$, where $a(t)$ is the cosmic
scale factor), and $D_A(z)$, the angular-diameter distance, which constrains $H(z)$ in an integral form.
The measurement of $H(z)$ allows us to determine the time dependence of dark energy.
The fact that we measure the redshifts of galaxies (and not their distances directly) leads to artifacts in the observed galaxy distribution,
the redshift space distortions (RSD). On large scales, the RSD are linear and enable the measurement of
the linear growth rate of cosmic large scale structure $f_g(z)$ \citep{Kaiser87}, which enables us to differentiate between dark energy
and modified gravity as the cause for cosmic acceleration, given the expansion history measurement \citep{Guzzo08,Wang08}.

The largest set of galaxy clustering data comes from the Baryon Oscillation Spectroscopic Survey (BOSS) 
[part of the Sloan Digital Sky Survey (SDSS) III]\footnote{http://www.sdss3.org/surveys/boss.php}, which should yield
millions of galaxy redshifts up to $z=0.7$ over 10,000 square degrees. BOSS has completed its observations in 2014.
The portfolio of ongoing and planned future galaxy redshift surveys is diverse and exciting.
The eBOSS survey\footnote{http://www.sdss.org/surveys/eboss/} (2014-2020) plans to cover over 7,500 square degrees
for luminous red galaxies (LRGs) in the redshift range of $0.6<z<0.8$, 
and over 1500 square degrees for [OII] emission line galaxies (ELGs) in the redshift range of $0.6<z<1$.
The DESI survey\footnote{http://desi.lbl.gov/} (2018-2022) will cover over 14,000 sq deg for LRGs ($0.1<z<1.1$) and [OII] ELGs ($0.1<z\la1.7$).
Euclid\footnote{http://www.euclid-ec.org/}, an ESA-led space mission scheduled for launch in 2020, 
will obtain galaxy redshifts for H$\alpha$ ELGs over 15,000 square degrees over a wide redshift range up to $z = 2$ \citep{RB}.
WFIRST\footnote{http://wfirst.gsfc.nasa.gov/} is NASA's next flagship mission in astrophysics, with a launch date in 2025. WFIRST is capable of a great range of possible galaxy redshift surveys
of H$\alpha$ and [OIII] ELGs, in the redshift range of 1 to 3; it will likely carry out a very deep galaxy redshift survey over at least 2000
square degrees that is complementary to the very wide galaxy redshift survey by Euclid \citep{SDT15}.

In order to fully realize the scientific potential of the ongoing and planned future surveys, it is important that we use BOSS data to
develop and test optimal approaches to extracting information on dark energy and modified gravity from galaxy clustering data.
Since the BOSS final data release (DR12) has not yet taken place, we use BOSS Data Release 10 (DR10) in this paper,
to explore the accurate modeling of small-scale galaxy clustering data in the context of the anisotropic analysis of the
two dimensional galaxy correlation function (2DCF). We use an MCMC-based model-independent approach to
measure $H(z)$, $D_A(z)$, and $f_g(z)\sigma_8(z)$ \citep{Song09} (with $\sigma_8(z)$ denoting the matter
power spectrum normalization on $8\,h^{-1}$Mpc scale at $z$), and marginalize over matter density $\Omega_m h^2$, baryon density $\Omega_b h^2$, 
power-law index of the primordial matter power spectrum $n_s$, normalization of the matter power spectrum today $P_0$, as well as parameters used
to model nonlinear effects and RSD. This conservative approach enables the combination of our results 
with other data to probe dark energy and gravity.

Our methodology is presented in Section~\ref{sec:method}. Our results are shown in Section~\ref{sec:results}.
We summarize and conclude in Section~\ref{sec:conclusion}.

\section{Methodology}
\label{sec:method}

\subsection{Modeling the Galaxy Correlation Function}
\label{subsec:model}

Our methodology is based on \cite{Wang14}, with the RSD modeling modified per \cite{Zheng13} (based on the work
of \cite{Zhang13}):
\be
\label{eq:Pk1}
P(\bfk)_{dw,nl}^{g,s}=b^2 P(\bfk)_{dw,nl}\left[1+\beta \tilde{W}(k,z)\mu^2\right]^2,
\ee
where $P(\bfk)_{dw,nl}^{g,s}$ is the redshift space galaxy power spectrum, $P(\bfk)_{dw,nl}$ is the matter power spectrum,
$b$ is the bias between galaxy and matter distributions, $\beta$ is the linear redshift space distortion parameter,
and $\mu$ is the cosine of the angle between $\bfk$ and the line-of-sight.
The window function $\tilde{W}(k,z)$ takes the form \citep{Zheng13}
\be
\tilde{W}(k,z)=\frac{1}{1+\Delta\alpha(z) \Delta^2(k,z)},
\ee
We find that it is simplest to choose  $\Delta^2(k,z)=k^3  P_{lin}/(2\pi^2)$, with the linear power spectrum given by
\be
\label{eq:dal}
P_{lin}=P_0 k^{n_s} T^2(k),
\ee
where $T(k)$ is the linear matter transfer function.

The nonlinear dwiggled matter power spectrum 
\be
P_{dw,nl}=F_{NL}(k)\,P_{dw,lin}(\bfk), 
\ee
with $F_{NL}(k)$ modeling nonlinear evolution and scale-dependent bias \citep{Cole05}:
\be
F_{NL}(k)=\frac{1+Qk^2}{1+f_Ak+Bk^2}.
\label{eq:NL}
\ee
We take $B=Q/10$ \citep{Sanchez08}.
We can write the linear dewiggled power spectrum as
\be
P_{dw,lin}(\bfk)= G^2(z)P_0 k^{n_s} \left\{T^2_{\rm nw}(k) + T^2_{\rm BAO}(k) e^{-g_\mu k^2 /(2k_*^2)}\right\},
\label{eq:P(k)dw}
\ee
where we have defined
\be
T^2_{\rm BAO}(k)=T^2(k)-T^2_{\rm nw}(k),
\ee
with $T_{\rm nw}(k)$ denoting the pure CDM (no baryons) transfer function given by
Eq.(29) from \cite{EH98}. The nonlinear damping factor, $e^{-g_\mu k^2 /(2k_*^2)}$, was derived using N-body simulations by \cite{Eisen07};
$g_\mu$ describes the enhanced damping along the line of sight due to the enhanced power:
\be
g_\mu(\bfk,z) \equiv G^2(z) \{ 1-\mu^2 +\mu^2 [1+f_g(z)]^2 \}.
\label{eq:gmu}
\ee
\noindent
Since density perturbations grow with cosmic time, the linear regime expands as we go to higher redshifts.
This is why the scale of the linear regime increases with $1/G(z)$ at high redshifts, while $g_\mu$ scales with the linear growth factor $G(z)$ squared.

The 2DCF, our model to be compare with data, is obtained by convolving $\tilde{\xi}$, the Fourier transform of 
the redshift space galaxy power spectrum $P(\bfk)_{dw,nl}^{g,s}$, with the probability distribution of galaxy peculiar velocities $f(v)$:
\be
\label{eq:model}
 \xi(\sigma,\pi)=\int_{-\infty}^\infty \tilde{\xi}\left(\sigma,\pi-\frac{v}{H(z)a(z)}
 \right)\,f(v)dv,
\ee
where $H(z)$ is the Hubble parameter and $a(z)$ is the cosmic scale factor,
and $f(v)$ is given by 
\be
 f(v)=\frac{1}{\sigma_v\sqrt{2\pi}}\exp\left(-\frac{v^2}{2\sigma_v^2}\right),
\ee
with $\sigma_v$ denoting the galaxy peculiar velocity dispersion. \cite{Zheng13} showed that this Gaussian $f(v)$ matches better
with their RSD modeling, compared to the usual form of $f(v)=(\sigma_v\sqrt{2})^{-1} \exp(-\sqrt{2}|v|/\sigma_v)$.

To save computational time in obtaining the Fourier transform of $P(\bfk)_{dw,nl}^{g,s}$, we write
\ba
\label{eq:Pk}
&&P(\bfk)_{dw,nl}^{g,s}=P(\bfk)_{nw,nl}^{g,s}+P(\bfk)_{BAO,dw,nl}^{g,s}\\
&&P(\bfk)_{nw,nl}^{g,s}=b^2 G^2(z)\left[1+\beta \tilde{W}(k) \mu^2\right]^2 P_0 k^{n_s} T^2_{\rm nw}(k) F_{NL}(k) \nonumber\\
&&P(\bfk)_{BAO,dw,nl}^{g,s}=b^2 G^2(z)\left[1+\beta  \tilde{W}(k) \mu^2\right]^2
P_0 k^{n_s} T^2_{\rm BAO}(k) \cdot \nonumber\\
& & \hskip 3cm \cdot F_{NL}(k) e^{-g_\mu k^2 /(2k_*^2)}\nonumber.
\ea
This leads to two terms in the Fourier transform of $P(\bfk)_{dw,nl}^{g,s}$ with different dependence on $\mu$:
\be
\tilde{\xi}(\sigma,\pi)=\xi^{g,s}_{nw}(\sigma,\pi)+\xi_{BAO,dw}^{g,s}(\sigma,\pi),
\ee
with $\sigma$ and $\pi$ denoting the transverse and line-of-sight separations of a pair of galaxies.
The second term is the Fourier transform of $P(\bfk)_{BAO,dw,nl}^{g,s}$, which is more complicated due to
the additional damping factor $e^{-g_\mu k^2 /(2k_*^2)}$, with $g_\mu$ dependent on $\mu$
(see Eq.[\ref{eq:gmu}]). \cite{CW13} found an easy way to deal with this by noting that the $\mu$-dependent damping factor in $k$-space 
becomes a Gaussian convolution in configuration space \citep{CW13}:
\be
 \label{eq:xi_gBAO}
 \xi^{g,s}_{BAO,dw}(\sigma,\pi)=\frac{1}{\sigma_\star\sqrt{\pi}}
\int_{-\infty}^\infty \rmd x\,\xi^{g,s}_{BAO,sdw}(\sigma,\pi-x)\, e^{-x^2/\sigma_\star^2},
\ee
where $\xi^{g,s}_{BAO,sdw}(\sigma,\pi)$ is the Fourier transform of $P(\bfk)_{BAO,dw,nl}^{g,s}$
with the damping factor $e^{-g_\mu k^2 /(2k_*^2)}$ replaced by its $\mu$-independent part,
$e^{-G^2(z) k^2 /(2k_*^2)}$, and 
\be
\sigma_\star^2=\frac{[4f_g(z)+2f^2_g(z)]G^2(z)}{k_\star^2}.
\ee

To calculate $\xi^{g,s}_{nw}(\sigma,\pi)$ and $\xi_{BAO,sdw}^{g,s}(\sigma,\pi)$, we take the Fourier transform of 
\be
P_x(\bfk)^s= P_x(k) \left[1+\beta \tilde{W}(k,z)\mu^2\right]^2.
\ee
This gives us
\ba
\tilde{\xi}^x(\sigma,\pi) &=& \xi^x(r) + 2 \beta \mu^2 \tilde{\xi}^x(r)  + \beta^2 \mu^4 \tilde{\xi}^x_2(r) \nonumber\\
&& + \frac{2}{3}\beta \left(1-3 \mu^2\right) \tilde{\overline{\xi}}^x(r) \nonumber \\
&& +\frac{\beta^2}{2} \left\{ \left(1-6\mu^2+5\mu^4\right)\tilde{\overline{\xi}}^x_2(r) \right.\nonumber\\
&& \left. -\frac{1}{5}\left(3-30\mu^2+35\mu^4\right)\tilde{\overline{\overline{\xi}}}^x_2(r) \right\}
\label{eq:xi}
\ea
where the superscript $x$ represents ``nw" or ``BAO,sdw". The function $\xi^x(r)$ is the Fourier transform of $P_x(k)$;
$\tilde{\xi}^x(r)$, $\tilde{\xi}^x_2(r)$, $\tilde{\overline{\xi}}^x(r)$, and $\tilde{\overline{\overline{\xi}}}^x_2(r)$
are related integrals that depend on the window function $\tilde{W}(k)$.
These are defined as follows:
\ba
&&\xi^x(r)=\frac{1}{2\pi^2}\int_0^\infty {\rm d} k\, k^2 P_x(k) \left[\frac{\sin(kr)}{kr}\right]\\
&&\tilde{\xi}^x(r)=\frac{1}{2\pi^2}\int_0^\infty {\rm d} k\, k^2 \tilde{W}(k) P_x(k) \left[\frac{\sin(kr)}{kr}\right]\\
&&\tilde{\xi}^x_2(r)=\frac{1}{2\pi^2}\int_0^\infty {\rm d} k\, k^2 \tilde{W}^2(k) P_x(k) \left[\frac{\sin(kr)}{kr}\right]\\
&&\tilde{\overline{\xi}}^x(r) = \frac{3}{r^3} \int_0^r {\rm d}s\, s^2 \tilde{\xi}^x(s)\\
&&\tilde{\overline{\xi}}_2^x(r) = \frac{3}{r^3} \int_0^r {\rm d}s\, s^2 \tilde{\xi}_2^x(s)\\
&&\tilde{\overline{\overline{\xi}}}_2^x(r) = \frac{5}{r^5} \int_0^r {\rm d}s\, s^4 \tilde{\xi}_2^x(s).
\label{eq:xidef}
\ea
Eqs.(\ref{eq:xi})-(\ref{eq:xidef}) give us $\xi^{g,s}_{nw}(\sigma,\pi)$ and $\xi_{BAO,sdw}^{g,s}(\sigma,\pi)$, with $P_x(k)$ given by 
\ba
&&P_{nw}(k)=b^2 G^2(z) P_0 k^{n_s} T^2_{\rm nw}(k) F_{NL}(k)\\
&&P_{BAO,sdw}(k)=b^2 G^2(z) P_0 k^{n_s} T^2_{\rm BAO}(k) F_{NL}(k) \cdot \nonumber\\
&& \hskip 3cm \cdot e^{-G^2(z) k^2 /(2k_*^2)} 
\ea
respectively. It is straightforward to check that Eqs.(\ref{eq:xi})-(\ref{eq:xidef}) give the standard expression for $\xi^s(\sigma, \pi)$
in terms of $P(\bfk)$ \citep{Hamilton92}, if we set $\tilde{W}=1$.

\subsection{Data and Covariance Matrix}
\label{subsec:data}

We use the publicly available CMASS sample from BOSS DR10 \citep{Anderson14}.
The DR10 CMASS sample consists of 540,147 galaxies over an effective area of 6161 deg$^2$, 
with 420,696 galaxies over an effective area of 4817 deg$^2$ in the Northern Galactic Cap, 
and 119,451 galaxies over 1345 deg$^2$ in the Southern Galactic Cap.
The CMASS sample is designed to be approximately stellar-mass-limited for $z>0.45$. 
The galaxies are color-selected, with a median redshift of 0.57.

The CMASS sample from DR10 has roughly twice the galaxy number and effective area compared to
the CMASS sample from DR9, which consists of 264,283 galaxies over an effective area of 3275 deg$^2$.
We used DR9 in \cite{Wang14}; it is appropriate for us to use DR10 in this paper to demonstrate our
improved modeling to extract small-scale cosmological information.

Mock catalogues are required to compute the covariance matrix for the data sample, and to validate our
analysis technique. We use the set of 600 mocks for BOSS DR10.
For a detailed description of these mocks\footnote{http://www.marcmanera.net/mocks/}, see \cite{Manera13} and \cite{Manera15}.
The input cosmological model of the mock catalogs is:
$\Lambda$CDM model with $\Omega_k=0$, $h=0.7$, $\Omega_m h^2=0.13426$ ($\Omega_m=0.274$), 
$\Omega_b h^2=0.0224$ ($\Omega_b=0.0457$), $n_s=0.95$, and $\sigma_8=0.8$.
We use this model as the fiducial model for our data analysis. 

Before carrying out our analysis of galaxy clustering, we need to convert measured redshifts of galaxies to comoving distances.
We use the fiducial model to make this conversion. Since our measurements of $H(z)$, $D_A(z)$, and $f_g(z)$ are made
through scaling (see Sec.\ref{sec:scaling}), our results are not sensitive to the assumed fiducial model.

To measure the 2DCF from data, we use the estimator \citep{Landy93}
\begin{equation}
\label{eq:xi_Landy}
\xi(\sigma,\pi) = \frac{DD(\sigma,\pi)-2DR(\sigma,\pi)+RR(\sigma,\pi)}{RR(\sigma,\pi)},
\end{equation}
where $\sigma$ and $\pi$ are the transverse and line-of-sight separations of a pair of galaxies in the sky.
DD, DR, and RR represent the normalized data-data, data-random, and random-random pair counts respectively in a given
distance range. The line-of-sight is defined as the direction from the observer to the 
center of a pair. We use a bin size of $8 \, h^{-1}$Mpc$\times 8 \,h^{-1}$Mpc. 
The estimator in Eq.(\ref{eq:xi_Landy}) has minimal variance for a Poisson
process. We use the random data sets that accompany the BOSS data sets,
which have the same radial and angular selection functions as the real data. 
To mitigate various systematic effects, the BOSS catalogs include weights that should be applied to each galaxy.

We calculate the 2DCF of the 600 mock catalogs, and use these to construct the covariance matrix of the 
measured 2DCF as follows:
\begin{equation}
 C_{ij}=\frac{1}{N-1}\sum^N_{k=1}(\bar{\xi}_i-\xi_i^k)(\bar{\xi}_j-\xi_j^k),
\label{eq:covmat}
\end{equation}
where $N$ is the number of the mock catalogs ($N=600$), $\bar{\xi}_m$ is the
mean of the $m^{th}$ bin of the mock catalog correlation functions, and
$\xi_m^k$ is the value in the $m^{th}$ bin of the $k^{th}$ mock catalog correlation function. 
To correct the under-estimate of the errors due to the finite number of mocks, we multiply the inverse covariance 
matrix by a factor of $(N-N_{data}-1)/(N-1)$, where $N_{data}$ is the number of data points used in our analysis \citep{Hartlap07}.

\subsection{The Likelihood Analysis}
\label{sec:scaling}

We follow the approach in \cite{CW12} and \cite{Wang14} in our likelihood analysis.
If the measurements are Gaussian distributed, the likelihood of 
a model given the data is proportional to $\exp(-\chi^2/2)$ \citep{Press92},
where $\chi^2$ compares data with model predictions.
We run Markov Chain Monte-Carlo (MCMC) \citep{Lewis02}, and assume the likelihood
${\cal L}\propto \exp(-\chi^2/2)$ in the acceptance function, with 
\be
 \label{eq:chi2}
 \chi^2\equiv\sum_{i,j=1}^{N_{bins}}\left[\xi_{th}(\bfs_i)-\xi_{obs}(\bfs_i)\right]
 C_{ij}^{-1}
 \left[\xi_{th}(\bfs_j)-\xi_{obs}(\bfs_j)\right]
\ee
where $\xi_{th}$ (see Sec.\ref{subsec:model}) and $\xi_{obs}$ 
(see Sec.\ref{subsec:data}) are the model and observed correlation functions
respectively. $N_{bins}$ is the number of data bins used, and $\bfs_i=(\sigma_i,\pi_i)$. 

For efficient and consistent implementation in the numerical analysis,
we avoid re-measuring the 2DCF from data for each model to obtain $\xi_{obs}$ in that model.
Instead, we use scaling to re-write Eq.(\ref{eq:chi2}), such that the model $\xi_{th}$ is
scaled in a consistent manner to be compared to the $\xi_{obs}$ measured assuming
the fiducial model. This works because the fiducial model is only used in converting redshifts into 
distances for the galaxies in our data sample; assuming different models in converting 
redshifts into distances results in observed galaxy distributions that are related by
a simple scaling of the galaxy separations. 
To derive this scaling, note that the separations of galaxies in angle and redshift are observables, 
thus independent of the model assumed, i.e.,
\ba
&&\Delta\theta = \frac{\sigma}{D_A(z)}=\frac{\sigma_{fid}}{D_A^{fid}(z)}\\
&& \Delta z= H(z) \pi = H^{fid}(z) \pi_{fid},
\ea
where the label ``fid" refers to parameters in the fiducial model, while the parameters without
the label represent an arbitrary model. For a thin redshift shell, we can now convert the galaxy 
separations from the fiducial model to another model using the scaling
(see, e.g., \cite{Seo03})
\be
\label{eq:scaling}
 (\sigma,\pi)=\left(\frac{D_A(z)}{D_A^{fid}(z)}\sigma_{fid},
 \frac{H^{fid}(z)}{H(z)}\pi_{fid}\right).
\ee
This means that the measured 2DCF's assuming an arbitrary model 
and the fiducial model are related as follows:
\be
\xi_{obs}(\sigma,\pi)= T\left(\xi^{fid}_{obs}(\sigma_{fid},\pi_{fid})\right),
\ee
with $T$ denoting the mapping given by Eq.(\ref{eq:scaling}).

Now the $\chi^2$ from Eq.(\ref{eq:chi2}) can be rewritten as \citep{CW12}
\ba 
\label{eq:chi2_2}
\chi^2 &\equiv&\sum_{i,j=1}^{N_{bins}}
 \left\{T^{-1}\left[\xi_{th}(\bfs_i)\right]-\xi^{fid}_{obs}(\bfs_i)\right\}
 C_{fid,ij}^{-1} \cdot \nonumber\\
 & & \cdot \left\{T^{-1}\left[\xi_{th}(\bfs_j)\right]-\xi_{obs}^{fid}(\bfs_j)\right\},
\ea
with $C_{fid}$ denoting the covariance matrix of the observed data 
assuming the fiducial model. The operator $T^{-1}\left[\xi_{th}(\bfs_i)\right]$
maps the model computed at $\{\sigma, \pi\}$ to the fiducial
model frame coordinates $(\sigma_{fid},\pi_{fid})$ as given by
Eq.(\ref{eq:scaling}).

We find that it is most efficient to convert the grid of $(\sigma_{fid},\pi_{fid})$
spanned by the measured 2DCF to a grid of $\{\sigma, \pi\}$ for each model
using Eq.(\ref{eq:scaling}), using the $H(z)$ and $D_A(z)$ assumed for that model. 
Then we compute the 2DCF for the model on the grid of $\{\sigma, \pi\}$, which depends on the other
parameters in the model: cosmological parameters $\Omega_m h^2$, $\Omega_b h^2$, $n_s$, $P_0$, 
as well as nonlinearity and RSD parameters $\beta$, $k_*$, $\Delta\alpha$, $f_g$, $\sigma_v$, $Q$, and $f_A$.
Finally, the model should be multiplied by a volume factor given by
\be
V_{fac}=\frac{H(z)}{H^{fid}(z)}\left(\frac{D_A^{fid}(z)}{D_A(z)}\right)^2.
\ee

Effectively, we are using the shape of the galaxy 2PCF as a standard ruler to measure $H(z)$ and $D_A(z)$, 
with cosmological parameters ($\Omega_m h^2$, $\Omega_b h^2$, $n_s$, $P_0$)
and parameters that describe systematic effects (nonlinearity and RSD) included as calibration parameters.
With reliable modeling of RSD, our technique also allows the measurement of $f_g(z)\sigma_8(z)$.

\section{Results}
\label{sec:results}

We have carried out the MCMC likelihood analysis of the BOSS DR10 CMASS sample, as well as a large number of the mocks.
The parameters that we have included are:
$H(0.57)$, $D_A(0.57)$, $\beta$, $\Omega_mh^2$, $\Omega_bh^2$, $n_s$, $P_{norm}$,  $\Delta\alpha$, $\sigma_v$, $k_\star$, $f_g(0.57)$, $Q$, and $f_A$.
The dimensionless normalization parameter $P_{norm}=P_0 b^2(0.57) G^2(0.57) [h\,$Mpc$^{-1}]^{n_s+3}$.

In post-processing of the MCMC chains, we also derive constraints on three key parameter combinations that 
are well constrained and insensitive to systematic effects:
\ba
&& x_h(0.57) \equiv H(0.57) r_s(z_d)/c\\
&& x_d(0.57) \equiv D_A(0.57)/r_s(z_d)\\
&& f_g(0.57) \sigma_8(0.57) = I_0^{1/2} P_{norm}^{1/2} \beta,
\ea
where we have defined
\be
I_0 \equiv  \int_0^\infty\rd \bar{k}\,  \frac{\bar{k}^{n_s+2}}{2\pi^2}\, 
T^2(\bar{k}\cdot h\mbox{Mpc}^{-1})\,
\left[\frac{3 j_1(8\bar{k})}{8\bar{k}}\right]^2,
\ee
where $\bar{k}\equiv k/[h\,\mbox{Mpc}^{-1}]$, 
and $j_1(kr)$ is spherical Bessel function. 
Note that the use of $\sigma_8$ does introduce an explicit $h$-dependence;
since $\sigma_8 \propto I_0=I_0(\Omega_m h^2, \Omega_b h^2, n_s, h)$; we compute $I_0$
with $h=0.7$ from the fiducial model. An alternative is to use $f_g(z)\sigma_m(z)$ as suggested by
\cite{WCH13}, with $\sigma_m(z)\equiv G(z) P_0h^3/$(Mpc)$^{3+n_s}$.
We have used $f_g(z)\sigma_8(z)$ here for comparison with the published results in the literature.
It is reassuring that the measured 2DCF does not depend on $h$, since
$k_{\parallel}$ and $k_\perp$ scale as $H(z)$ and $1/D_A(z)$ respectively \citep{WCH13}.

To facilitate easy comparison between data and models, we define the comoving sound horizon 
at the drag epoch $z_d$ as given by
\ba
\label{eq:rs}
r_s(z_d)  &= & \int_0^{t} \frac{c_s\, dt'}{a}
=cH_0^{-1}\int_{z}^{\infty} dz'\,
\frac{c_s}{E(z')}, \\
 &= & cH_0^{-1} \int_0^{a} 
\frac{da'}{\sqrt{ 3(1+ \overline{R_b}\,a')\, {a'}^4 E^2(z')}}\nonumber\\
&=& \frac{2997.9\,\mbox{Mpc}}{\sqrt{0.75 \overline{R_b}\omega_m}}\,
\ln\left\{ \frac{\sqrt{a_d+a_{eq}}+\sqrt{a_d+\overline{R_b}^{-1}}}
{\sqrt{a_{eq}}+\sqrt{\overline{R_b}^{-1}}}
\right\}, \nonumber
\ea
where $a$ is the cosmic scale factor, $a =1/(1+z)$;
$a^4 E^2(z)=\Omega_m (a+a_{\rm eq})+\Omega_k a^2 +\Omega_X X(z) a^4$,
with $a_{\rm eq}=\Omega_{\rm rad}/\Omega_m=1/(1+z_{\rm eq})$, and
$z_{\rm eq}=2.5\times 10^4 \Omega_m h^2 (T_{CMB}/2.7\,{\rm K})^{-4}$.
The sound speed is $c_s=1/\sqrt{3(1+\overline{R_b}\,a)}$,
with $\overline{R_b}\,a=3\rho_b/(4\rho_\gamma)$,
$\overline{R_b}=31500\Omega_bh^2(T_{CMB}/2.7\,{\rm K})^{-4}$.
We take $T_{CMB}=2.72548$ \citep{Fixsen09}.
We assume the redshift of the drag epoch $z_d$ to be \citep{EH98}
\begin{equation}
z_d  =
 \frac{1291(\Omega_mh^2)^{0.251}}{1+0.659(\Omega_mh^2)^{0.828}}
\left[1+b_1(\Omega_bh^2)^{b2}\right],
\label{eq:zd}
\end{equation}
with
\begin{eqnarray}
  b_1 &= &0.313(\Omega_mh^2)^{-0.419}\left[1+0.607(\Omega_mh^2)^{0.674}\right], \nonumber\\
  b_2 &= &0.238(\Omega_mh^2)^{0.223}.
  \label{eq:zd_b}
\end{eqnarray}

Our choice for $r_s(z_d)$ differs from that of the BOSS team, who have chosen to define
$r_s(z_d)$ as the value computed numerically by CAMB.
For a given cosmological model, our $r_s(z_d)$ value from Eqs.(\ref{eq:rs})-(\ref{eq:zd_b})
differs from that given by CAMB by a factor which is close to one and 
nearly independent of the cosmological model \citep{Mehta12}.
Since $r_s(z_d)$ is only used to scale $H(z)$ and $D_A(z)$,
the comparison between data and models should be insensitive to the choice of $r_s(z_d)$,
as long as we are consistent in using the same definition of $r_s(z_d)$ in analyzing data
and making model predictions.

We apply flat priors on all the parameters. The priors on the parameters that are well constrained
by the data, $H(0.57)$, $D_A(0.57)$, $\beta$, $\Omega_m h^2$, $P_{norm}$, $\Delta\alpha$, 
are sufficiently wide so that the results are insensitive to the ranges chosen.
We impose flat priors of $\Omega_bh^2=(0.02018, 0.02438)$,  $n_s=(0.9137, 1.0187)$,
corresponding to the 7$\sigma$ range of these parameters from the first year
Planck data, with $\sigma$ from the Gaussian fits by \cite{Wang13};
these wide priors ensure that CMB constraints are not double counted 
when our results are combined with CMB data \citep{CWH12}.
Our results are not sensitive to the parameters that describe the systematic
uncertainties, $k_\star$, $f_g(0.57)$, $\sigma_v$, $Q$, $A$; we impose reasonable
flat priors on these: $k_\star=(0.1, 0.3)$, $f_g(0.57)=0.35-0.55$, $\sigma_v<500\,$km/s, $Q=0-40\,($Mpc$/h)^2$, and $f_A=0-10\,$Mpc$/h$.

\subsection{Validation Using Mocks}

\begin{figure}
\centering
\includegraphics[width=0.83\columnwidth,clip]{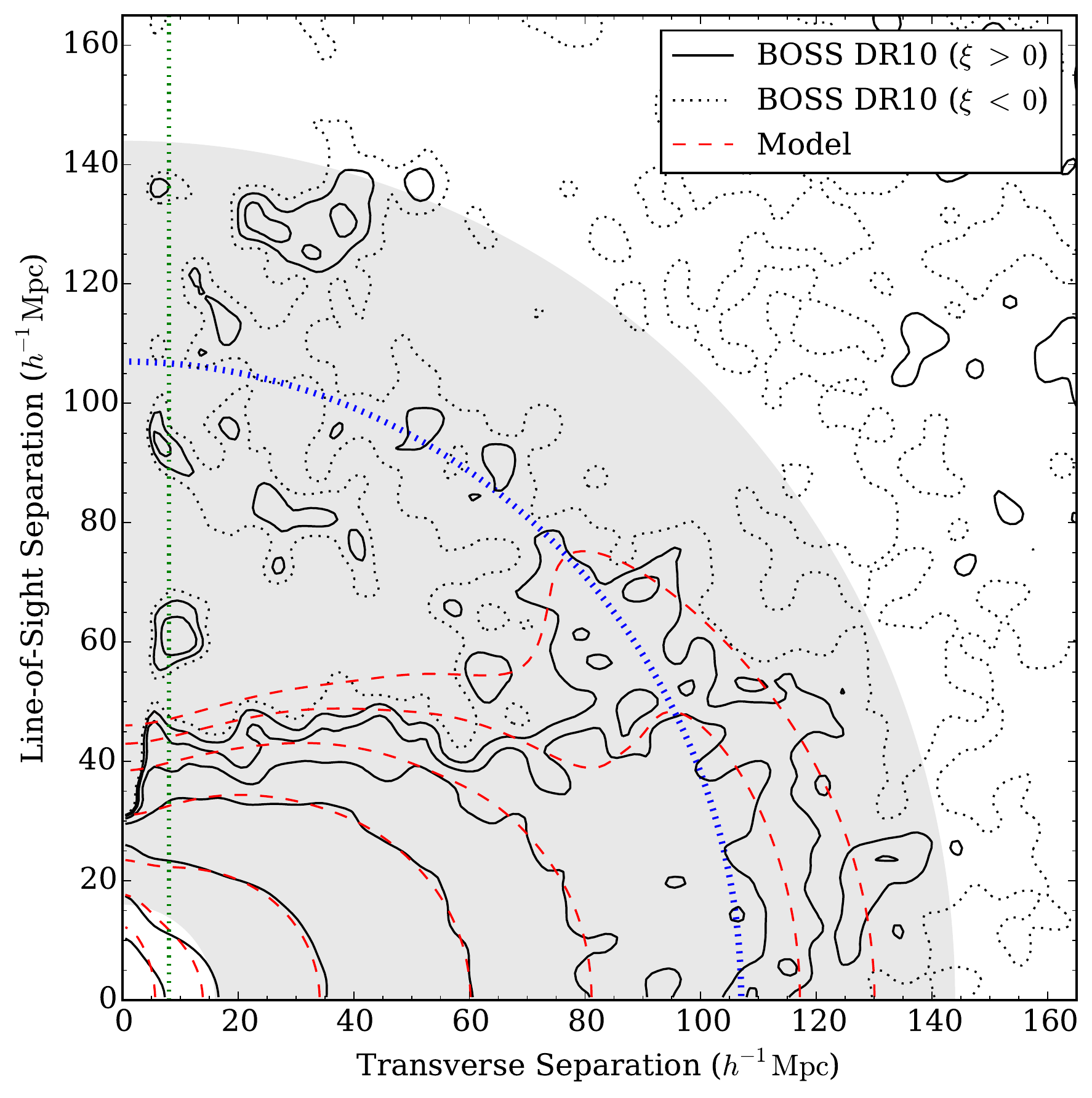}
\includegraphics[width=0.83\columnwidth,clip]{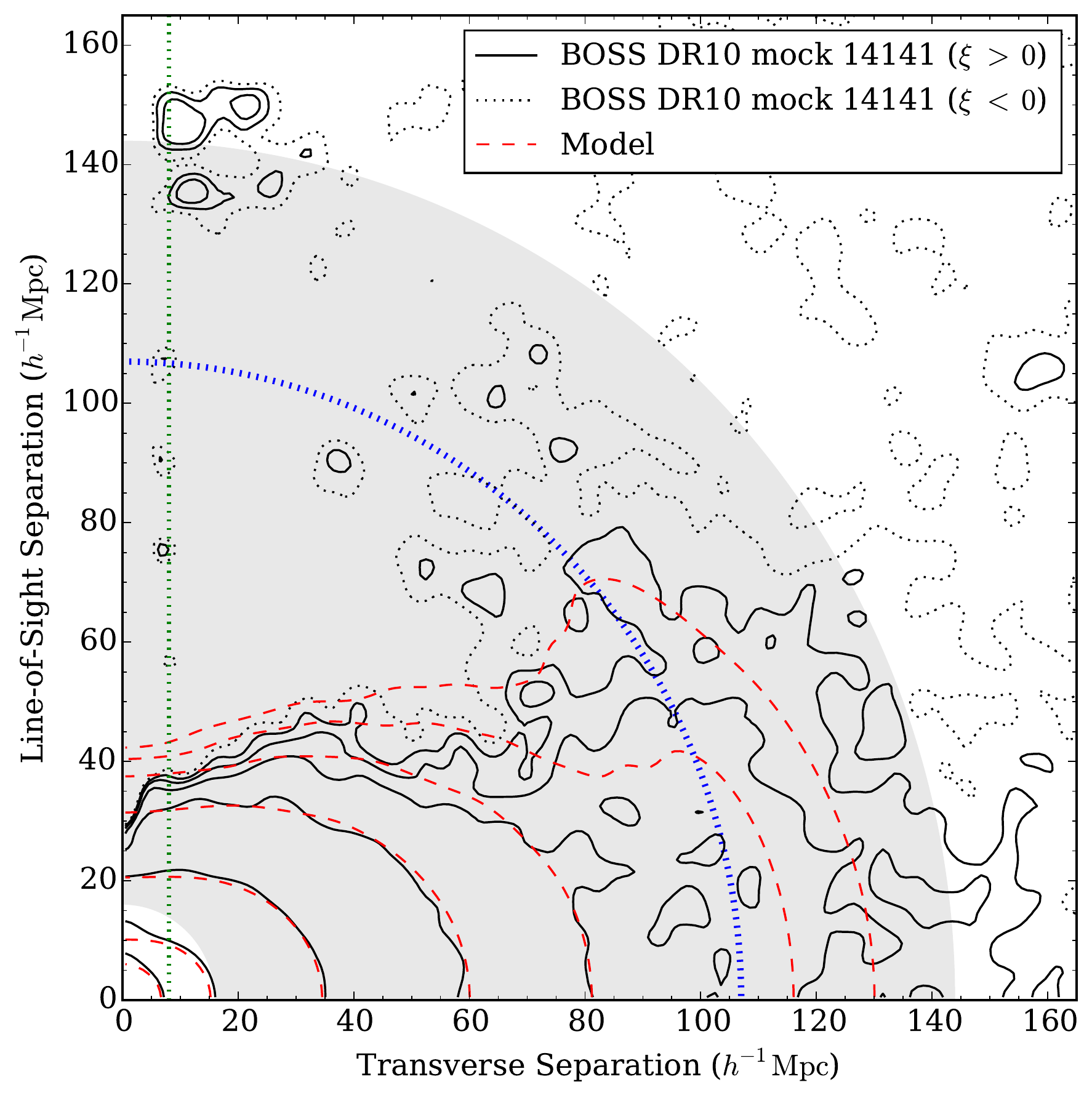}
\caption{The BOSS DR10 CMASS sample (upper panel) and a representative mock (lower panel).
The contour levels are $\xi=0.0025,0.005,0.01,0.025,0.1,0.5,2.0$,
with the dotted contours denoting $\xi \leq 0$. 
The solid lines are the data (or mock data); the dashed lines are our best-fit model.}
\label{fig:xi2d}
\end{figure}
Figure \ref{fig:xi2d} shows the BOSS DR10 CMASS sample (upper panel) and a representative mock (lower panel).
The contour levels are $\xi=0.0025,0.005,0.01,0.025,0.1,0.5,2.0$; the dotted contours denote
$\xi \leq 0$. The solid lines are the data (or mock data), and the dashed lines 
are our best-fit model. The comparison of Figure \ref{fig:xi2d} (BOSS DR10) with Fig.1 in \cite{Wang14} (BOSS DR9) shows the significant
expansion in the range over which the 2DCF from data is well determined. The bottom panel in Figure \ref{fig:xi2d} clearly shows that our
model applies even on small scales. The shaded disk indicates the range of scales that we will use in our MCMC likelihood analysis to
measure $H(z)$, $D_A(z)$, and $f_g(z)$, $16-144\,h^{-1}$Mpc.

We have analyzed 264 mocks of the BOSS DR10 CMASS sample using MCMC likelihood analysis, in the scale range of $16-144\,h^{-1}$Mpc.
To speed up computation, we fixed the nonlinearity parameters $Q$ and $f_A$ to fiducial values of $Q=13$ and $f_A=1.5$.
We find that including the data at $\sigma < 8\,h^{-1}$Mpc leads to high noise levels, and results in $f_g(0.57)\sigma_8(0.57)$ measurements 
that are biased high compared to the true value.
However, discarding the data at $\sigma < 8\,h^{-1}$Mpc leads to $H(0.57)$ measurements that are biased low compared to the true value.
The data contours (upper panel in Figure \ref{fig:xi2d}) suggest that we discard the data at $\sigma < 8\,h^{-1}$Mpc for $\pi>48\,h^{-1}$Mpc only,
so that we can use the less noisy data near the line of sight on intermediate scales.
We find that this cut leads to unbiased estimates of
$x_h=H(0.57) \,r_s(z_d)/c$, $x_d=D_A(0.57)/r_s(z_d)$, and $f_g(0.57)\sigma_8(0.57)$. 

Fig.{\ref{fig:meanxhxd} presents the resultant likelihood peak distributions of $x_h(0.57)$, $x_d(0.57)$, and $f_g(0.57)\sigma_8(0.57)$ from 264 mocks
with the scale range of $16-144\,h^{-1}$Mpc (solid lines), and 252 mocks with the scale range of $32-144\,h^{-1}$Mpc (dashed lines).
These show the distributions of the best-fit values from the mocks. The dotted lines indicate the values predicted by the input model of the mocks. 
The true values of $x_h$, $x_d$, and $f_g\sigma_8$ are all near the mean values in the distributions of the best-fit values for the scale range
of $16-144\,h^{-1}$Mpc, but are somewhat farther away from the mean values for the scale range of $32-144\,h^{-1}$Mpc. 
This indicates that our modeling works remarkably well for the scale range of $16-144\,h^{-1}$Mpc, giving unbiased parameter estimates.
For the scale range of $32-144\,h^{-1}$Mpc, the parameter estimates are slightly biased.
Comparing our Fig.{\ref{fig:meanxhxd} with Fig.2 of \cite{Wang14}, one can 
see that our current modeling significantly improves the recovery of the true $H(0.57)$.

Note that we have plotted the best-fit values, and not the marginalized means, of $x_h(0.57)$, $x_d(0.57)$, and $f_g(0.57)\sigma_8(0.57)$ from the mocks.
This is because the best-fit values are obtained much more quickly than the converged marginalized means (which are sensitive to the tails of the 
distributions). As the MCMC chains converge, the marginalized means approach the likelihood peak (i.e., the best-fit) values, and the two become 
very similar \citep{Lewis02}.

\begin{figure}
\centering
\includegraphics[width=0.65\columnwidth,clip]{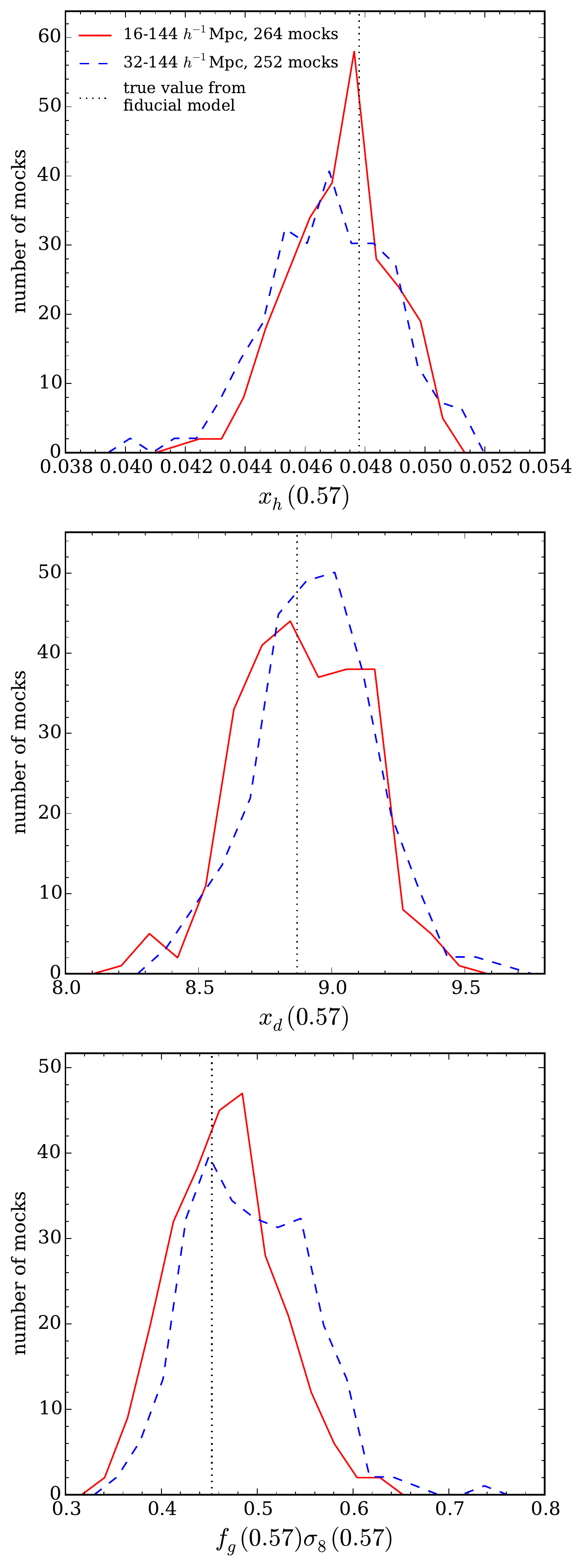}
\caption{The distribution of best-fit values of $x_h(0.57)=H(0.57) \,r_s(z_d)/c$, $x_d(0.57)=D_A(0.57)/r_s(z_d)$, and $f_g(0.57)\sigma_8(0.57)$ from 264 mocks 
of the BOSS DR10 CMASS sample with the scale range of $16-144\,h^{-1}$Mpc (solid lines), and 252 mocks with the scale range of $32-144\,h^{-1}$Mpc (dashed lines). 
The dotted lines indicate the values predicted by the input model of the mocks.}
\label{fig:meanxhxd}
\end{figure}

\subsection{Results from BOSS DR10 CMASS Sample}

We now present our results from analyzing the real data, the BOSS DR10 CMASS sample.
We use the same methodology as we have used for the mocks.
Table \ref{table:chi2} lists the $\chi^2$ per degree of freedom from the different cases that we have studied. The ``$\sigma$ \& $\pi$ cut" refers to
excluding the narrow wedge along the line-of-sight at $\sigma< 8\,h^{-1}$Mpc for $\pi>48\,h^{-1}$Mpc, the same cut as we used for the mocks.
All four cases with the $\sigma$ \& $\pi$ cut have $\chi^2_{pdf} \simeq 1$, while the no $\sigma$ \& $\pi$ cut case has $\chi^2_{pdf} \simeq 1.5$;
this supports our choice of making the $\sigma$ \& $\pi$ cut in the remainder of our analysis.
\begin{table*}
\begin{center}
\begin{tabular}{lcccccccc}\hline
scale range & $\sigma$ \& $\pi$ cut & $\Delta\alpha$ & $(Q, f_A)$ & $N_{data}$ & $N_{par}$ & $\chi^2_{min}$ & $\chi^2_{pdf}$ & comment\\ \hline
16-144$\,h^{-1}$Mpc & Yes &Varied & $(13, 1.5)$ & 240 & 11 & 248.5 & 1.09 & validated by mocks\\
\hline
32-144$\,h^{-1}$Mpc & Yes &Varied & $(13, 1.5)$ & 230 & 11 & 209.0 & 0.95 & high $f_g(0.57)\sigma_8(0.57)$\\
16-144$\,h^{-1}$Mpc & No & Varied & $(13, 1.5)$ & 252 & 11 & 355.0 & 1.47 & high $f_g(0.57)\sigma_8(0.57)$ \\
16-144$\,h^{-1}$Mpc & Yes & Zero & $(13, 1.5)$ & 240 & 10 & 255.7 & 1.11 & low $f_g(0.57)\sigma_8(0.57)$\\
16-144$\,h^{-1}$Mpc & Yes & Varied & Varied & 240 & 13 & 244.7 & 1.08 & slow convergence\\
\hline
\end{tabular}
\end{center}
\caption{$\chi^2$ per degree of freedom in the modeling of the BOSS DR10 CMASS sample, for different data selection and modeling choices.} 
\label{table:chi2}
\end{table*}

Fig.\ref{fig:params_pdf} shows the 1D marginalized probability distribution of parameters measured from BOSS DR10 CMASS sample,
for the four cases in Table \ref{table:chi2} with the $\sigma$ \& $\pi$ cut.
The solid lines are results for the scale range $16-144\,h^{-1}$Mpc, with nonlinearity parameters $Q=13$ and $f_A=1.5$.
The dashed lines show what happens if we vary $Q$ and $f_A$:
the constraints on $H(0.57) \,r_s(z_d)/c$, $D_A(0.57)/r_s(z_d)$,  and $f_g(0.57)\sigma_8(0.57)$ remain essentially unchanged.
The slight differences are due to the MCMC chains with varying $Q$ and $f_A$ not having fully converged; these are very slow to converge
due to the weak constraints on $Q$ and $f_A$ from data.

The dot-dashed lines in Fig.\ref{fig:params_pdf} show the results of choosing a narrower scale range that leaves out the smallest scale information: $32-144\,h^{-1}$Mpc.
We find that not using the small scale information from $16-32\,h^{-1}$Mpc
leads to a much weaker constraint on $H(0.57) \,r_s(z_d)/c$, and higher values for $f_g(0.57)\sigma_8(0.57)$, while having only a minor impact on
the constraints on $D_A(0.57)/r_s(z_d)$.
It is surprising that the scale ranges $16-144\,h^{-1}$Mpc and $32-144\,h^{-1}$Mpc give significantly different constraints on 
$H(0.57)$, $D_A(0.57)$, and $\Omega_m h^2$; this suggests that there are significant degeneracies in fitting the model to the data, with the
addition of the small scale data breaking the degeneracy. It is reassuring that the two scale ranges give similar constraints on the
physical parameters  $H(0.57) \,r_s(z_d)/c$ and $D_A(0.57)/r_s(z_d)$, in agreement with the results from the mocks (see Fig. \ref{fig:meanxhxd}).

The dotted lines in Fig.\ref{fig:params_pdf} show the results from setting $\Delta\alpha=0$, i.e., not using the RSD modeling from \cite{Zheng13},
for the scale range of $16-144\,h^{-1}$Mpc.
This has a minimal impact on the $H(0.57) \,r_s(z_d)/c$ measurement, but significantly weakens the $D_A(0.57)/r_s(z_d)$ measurement,
and leads to very low values for $f_g(0.57)\sigma_8(0.57)$. This is not surprising; the measurement of the growth rate
is highly sensitive to the modeling of RSD on small scales.

\begin{figure}
\vspace{-0.2in}
\hskip -0.6in
\includegraphics[width=1.3\columnwidth,clip]{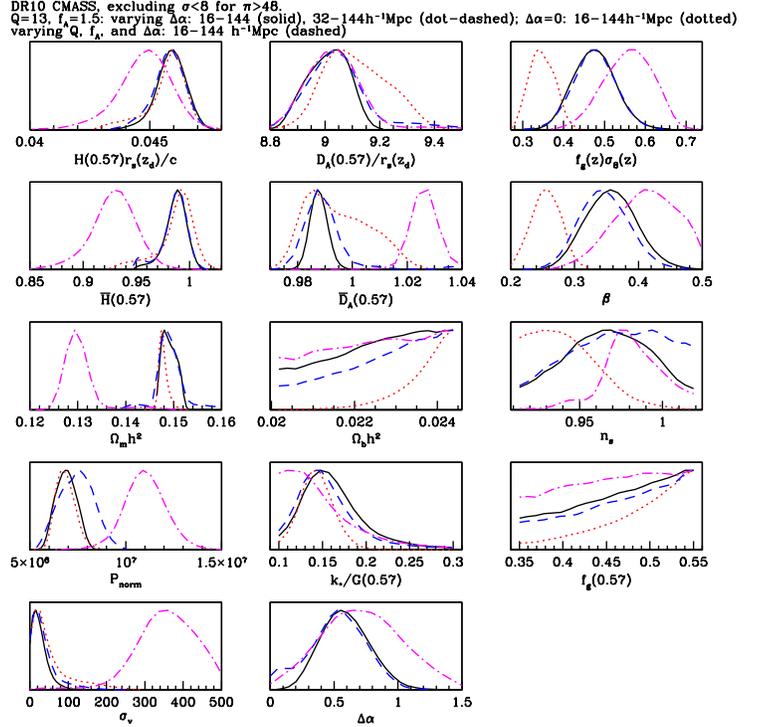}
\caption{The 1D marginalized probability distribution of parameters 
estimated from the BOSS DR10 CMASS sample, excluding $\sigma<8\,h^{-1}$Mpc for $\pi>48\,h^{-1}$Mpc. 
The different line types represent different choices made in our analysis (all are tabulated in Table \ref{table:chi2}).
The solid lines are for the scale range of $16-144h^{-1}$Mpc, varying $\Delta\alpha$, $Q=13$, $f_A=1.5$.
The dot-dashed lines are for the scale range of $32-144h^{-1}$Mpc, varying $\Delta\alpha$, $Q=13$, $f_A=1.5$.
The dotted lines are for the scale range of $16-144h^{-1}$Mpc, $\Delta\alpha=0$, $Q=13$, $f_A=1.5$.
The dashed lines are for the scale range of $16-144h^{-1}$Mpc, varying $\Delta\alpha$, $Q$, and $f_A$.}
\label{fig:params_pdf}
\end{figure}

Table \ref{table:means} gives the marginalized means and standard deviations of $\{H(0.57)$, $D_A(0.57)$, $\Omega_m h^2$, $\beta$,  $H(0.57) \,r_s(z_d)/c$, $D_A(0.57)/r_s(z_d)$, 
$f(0.57)\,\sigma_8(0.57)\}$ from BOSS DR10 CMASS sample, for the scale ranges $16<s<144h^{-1}$Mpc and $32<s<144h^{-1}$Mpc (excluding $\sigma<8 \,h^{-1}$Mpc for $\pi>48\,h^{-1}$Mpc).
The differences between the constraints on $\{H(0.57) \,r_s(z_d)/c$, $D_A(0.57)/r_s(z_d)$,  $f(0.57)\,\sigma_8(0.57)\}$ for the two different scale ranges 
are in qualitative agreement with that found using mocks (see Fig.\ref{fig:meanxhxd}).
Since the mocks show that the $f(0.57)\,\sigma_8(0.57)$ measurements from $16<s<144h^{-1}$Mpc are unbiased, we draw the same conclusion about the measurements from the real data.
This implies that the $f(0.57)\,\sigma_8(0.57)$ measurement from $32<s<144h^{-1}$Mpc from the real data is biased high.
Note that while the measurements of $f(0.57)\,\sigma_8(0.57)$ differ significantly for the two scale ranges, they overlap at 1$\sigma$, indicating
that the difference is statistically consistent with the predictions from the mocks.
Table \ref{table:covmat} shows the corresponding normalized covariance matrix for the case with the validated scale range of $16<s<144h^{-1}$Mpc.
\begin{table}
\begin{center}
\begin{tabular}{cll}\hline
&$16<s<144$ &$32<s<144$ \\ \hline
$	H(0.57)	$& 92.36 $\pm$ 0.98 & 86.95$\pm$1.96\\
$	D_A(0.57)$& 1343.00 $\pm$ 4.26 & 1395.51$\pm$7.85\\
$	\Omega_m h^2$& 0.1492$\pm$  0.0014&  0.1304 $\pm$ 0.0037\\
$	\beta	$&  0.358 $\pm$  0.039  &  0.412$\pm$  0.048\\
\hline
$	H(0.57) \,r_s(z_d)/c	$& 0.0459 $\pm$ 0.0006 & 0.0448$\pm$  0.0011\\ 
$	D_A(0.57)/r_s(z_d)	$& 9.0107 $\pm$  0.0729 & 9.0353 $\pm$  0.1050\\
$	f(0.57)\,\sigma_8(0.57)	$& 0.4757 $\pm$  0.0497 & 0.5583 $\pm$ 0.0579\\
\hline

\end{tabular}
\end{center}
\caption{
The mean and standard deviation of 
$\{H(0.57)$, $D_A(0.57)$, $\Omega_m h^2$, $\beta$,  $H(0.57) \,r_s(z_d)/c$, $D_A(0.57)/r_s(z_d)$, 
$f(0.57)\,\sigma_8(0.57)\}$ from BOSS DR10 CMASS sample, for the scale ranges $16<s<144h^{-1}$Mpc and $32<s<144h^{-1}$Mpc
(excluding $\sigma<8 \,h^{-1}$Mpc for $\pi>48\,h^{-1}$Mpc). The unit of $H$ is $\Hunit$. The unit of $D_A$ is $\rm Mpc$.
} \label{table:means}
\end{table}

\begin{table*}
\begin{center} 
\begin{tabular}{crrrrrrrr}\hline
&  $H(0.57)$ &$D_A(0.57)$ &$\Omega_m h^2$ & $\beta$ &$H(0.57) \,r_s(z_d)/c$ & $D_A(0.57)/r_s(z_d)$ & $f(0.57)\sigma_8(0.57)$ \\
 $H(0.57)$ &      1.0000    &  0.3191  &   $-$0.0092   &   0.1748 &     0.8239  &    0.0879     & 0.1104\\
$D_A(0.57)$&       0.3191     & 1.0000   &  $-$0.0983   &   0.0899   &   0.2592   &   0.3856 &    0.1016\\
$\Omega_mh^2$&     $-$0.0092   &  $-$0.0983  &    1.0000  &    0.0243  &   $-$0.2370   &   0.3364  &    0.0411\\
$\beta$&      0.1748    &  0.0899   &   0.0243 &     1.0000 &     0.1558  &   0.0098  &    0.9845\\
$H(0.57) \,r_s(z_d)/c$&      0.8239 &    0.2592  &   $-$0.2370   &   0.1558  &    1.0000    & $-$0.4514  &    0.1115\\
$D_A(0.57)/r_s(z_d)$&       0.0879   &   0.3856  &    0.3364  &    0.0098 &    $-$0.4514  &    1.0000   &   0.0025\\
$f(0.57)\sigma_8(0.57)$&       0.1104   &   0.1016  &    0.0411 &     0.9845   &   0.1115  &    0.0025   &   1.0000\\
\hline
\end{tabular}
\end{center}
\caption{Normalized covariance matrix of the measured and derived parameters, $\{H(0.57)$, $D_A(0.57)$, $\Omega_m h^2$, $\beta,$, 
$H(0.57) \,r_s(z_d)/c$, $D_A(0.57)/r_s(z_d)$, $f(0.57)\sigma_8(0.57)\}$,  from the BOSS DR10 CMASS sample for the scale range of $16<s<144h^{-1}$Mpc.}
 \label{table:covmat}
\end{table*}

\section{Summary and Discussion}
\label{sec:conclusion}

Galaxy clustering is a key probe of dark energy and modified gravity. Much of its ultimate power will come from small-scales,
which can only be included in the data analysis if we can reliably model galaxy clustering on these scales.
We have presented a new approach to measuring cosmic expansion history and growth rate of large scale structure using
the anisotropic two dimensional galaxy correlation function (2DCF) measured from data over the wide scale range
of 16-144$\,h^{-1}$Mpc, reaching down to a significantly smaller scale than in previous work. Our modeling of galaxy clustering uses
the empirical modeling of small-scale galaxy clustering derived from numerical simulations by Zheng et al. (2013) (see Eqs.[\ref{eq:Pk1}]-[\ref{eq:dal}]), 
which provides improved fit to RSD and nonlinear effects on small scales.
We have validated our methodology using mock catalogues, finding it to enable accurate and precise measurements of 
cosmic expansion history and growth rate of large scale structure.

Applying our methodology to the analysis of the 2DCF of galaxies from the BOSS DR10 CMASS sample, 
in the scale range of 16 to 144 $h^{-1}$Mpc (excluding the noisy data in the small line-of-sight wedge beyond 48 $h^{-1}$Mpc),
we measure $H(0.57)r_s(z_d)/c$, $D_A(0.57)/r_s(z_d)$, and $f_g(0.57)\sigma_8(0.57)$ with precisions of 1.3\%, 0.8\%, 
and 10.5\% respectively (see Table \ref{table:means}).
These are significantly tighter than those obtained by others using the same data, see e.g., \cite{Anderson14}.
This is not surprising, since we have utilized significantly more information from data.

It is often assumed that discarding small-scale information leads to more robust measurements of $H(z)$ and $D_A(z)$.
We find that neglecting the small-scale information weakens the constraints on $H(z)r_s(z_d)$ and $D_A(z)/r_s(z_d)$,
as expected (see Fig.\ref{fig:params_pdf}). Interestingly, omitting the small-scale
information seems to favor a low matter density, along with a low $H(z)$ and a high $D_A(z)$, which combine to give 
roughly the same $H(z)r_s(z_d)$ and $D_A(z)/r_s(z_d)$ but with larger uncertainties, compared to including the small-scale
information. This indicates that the measurements of $H(z)r_s(z_d)$ and $D_A(z)/r_s(z_d)$ are more robust than that of $H(z)$ and $D_A(z)$.

We find that the measurement of $f_g(z)\sigma_8(z)$ is very sensitive to the RSD modeling.
Not including the improved RSD modeling from \cite{Zheng13} leads to an estimate of $f_g(z)\sigma_8(z)$ that is
biased low significantly (see Fig.\ref{fig:params_pdf}). 
On the other hand, omitting the small-scale information, even when using the RSD modeling from \cite{Zheng13}, 
leads to an estimate of $f_g(z)\sigma_8(z)$ that is biased somewhat high. Our conclusion that the $f_g(z)\sigma_8(z)$
measurement from $32<s<144h^{-1}$Mpc is biased high (while that from $16<s<144h^{-1}$Mpc is unbased) is based on tests using the mocks (see Fig.\ref{fig:meanxhxd}).
The trends discussed above may explain in part the wide range of $f_g(0.57)\sigma_8(0.57)$ measurements from BOSS data that have been reported in the literature.

It is surprising that using the data in the scales ranges of $16<s<144h^{-1}$Mpc and $32<s<144h^{-1}$Mpc give very different constraints on
$H(0.57)$, $D_A(0.57)$, and $\Omega_m h^2$ (see Table \ref{table:means}). These two scale ranges do give similar constraints for the physical parameters
$H(0.57)r_s(z_d)/c$, $D_A(0.57)/r_s(z_d)$, with the differences in qualitative agreement with the results from mocks (see Fig.\ref{fig:meanxhxd}).
This suggests that the different constraints on $H(0.57)$, $D_A(0.57)$, and $\Omega_m h^2$ 
result from degeneracies in fitting the model to the data; the addition of the small scale data breaks this degeneracy. 
This indicates that $H(0.57)r_s(z_d)/c$ and $D_A(0.57)/r_s(z_d)$, instead of $H(0.57)$, $D_A(0.57)$, and $\Omega_m h^2$,
should be used to summary BAO constraints.

Another surprise may be how well our model fits, since we used the model from \cite{Zheng13} (based on \cite{Zhang13}),
which is similar to the model proposed by \cite{Sco04}, which is not expected to be accurate beyond $k=0.1\,h$Mpc$^{-1}$, or
a scale of $40-50\,h^{-1}$Mpc. The difference between \cite{Sco04} and \cite{Zhang13} is that the earlier work did not
explicitly make the RSD model corrections a modification to the linear model by \cite{Kaiser87} in the form of a window function.
The introduction of the window function by \cite{Zhang13} allows a compact formulation for the RSD model that is easily
implemented in the framework from \cite{Wang14}, which already includes a correction factor for nonlinear evolution and
scale-dependent bias (see Eq.[\ref{eq:NL}]), as well as the dewiggled power spectrum (see Eqs.[\ref{eq:P(k)dw}]-[\ref{eq:gmu}]), with asymmetric damping that accounts
for the damping of the BAO peak due to nonlinear effects. Our new model, presented in this paper, combines these three models,
with the parameters in each determined by data. This proves adequate for fitting the BOSS DR10 data.

We have not included massive neutrinos in our analysis, since they would likely have a small effect, and are computationally expensive.
However, it is important to include massive neutrinos in data analysis; we will do so in future work.

Our results are encouraging, and indicate that we can significantly tighten constraints on dark energy and modified gravity by reliably
modeling small-scale galaxy clustering. We will apply our methodology to BOSS DR12 data, once they are publicly available.
We will also include this new approach in the forecasting of constraints on dark energy and gravity for Euclid and WFIRST.

\section*{Acknowledgments}

The analysis of the mocks were carried out on the supercomputing clusters at Jet Propulsion Laboratory.
I am grateful to Chia-Hsun Chuang for sharing the 2DCF of the BOSS DR10 CMASS sample and the 600 mocks,
and Alex Merson for helping me make Fig.1 and Fig.2 using python.

Funding for SDSS-III has been provided by the Alfred P. Sloan Foundation, the Participating Institutions, the National Science Foundation, and the U.S. Department of Energy Office of Science. The SDSS-III web site is http://www.sdss3.org/.

SDSS-III is managed by the Astrophysical Research Consortium for the Participating Institutions of the SDSS-III Collaboration including the University of Arizona, the Brazilian Participation Group, Brookhaven National Laboratory, Carnegie Mellon University, University of Florida, the French Participation Group, the German Participation Group, Harvard University, the Instituto de Astrofisica de Canarias, the Michigan State/Notre Dame/JINA Participation Group, Johns Hopkins University, Lawrence Berkeley National Laboratory, Max Planck Institute for Astrophysics, Max Planck Institute for Extraterrestrial Physics, New Mexico State University, New York University, Ohio State University, Pennsylvania State University, University of Portsmouth, Princeton University, the Spanish Participation Group, University of Tokyo, University of Utah, Vanderbilt University, University of Virginia, University of Washington, and Yale University.


\setlength{\bibhang}{2.0em}

\label{lastpage}


\begin{thebibliography}{}

  \setlength{\itemindent}{-2.5em}


\bibitem[Anderson et al.(2014)]{Anderson14}
Anderson, L., et al., 2014, MNRAS, 441, 24

\bibitem[Blake \& Glazebrook(2003)]{Blake03}
Blake C., Glazebrook G. 2003, ApJ 594, 665

\bibitem[Caldwell \& Kamionkowski(2009)]{Caldwell09}
Caldwell, R. R., \& Kamionkowski, M., 2009, Ann.Rev.Nucl.Part.Sci., 59, 397

\bibitem[Chuang \& Wang(2012)]{CW12}
Chuang C.-H., Wang Y. 2012, MNRAS 426, 226

\bibitem[Chuang \& Wang(2013)]{CW13}
Chuang C.-H., Wang Y. 2013, MNRAS, 435, 255

\bibitem[Chuang, Wang, \& Hemantha(2012)]{CWH12}
Chuang, C.-H., Wang Y., Hemantha M. 2012, MNRAS 423, 1474

\bibitem[Chuang et al.(2013)]{Chuang13}
Chuang, C.-H., et al. 2013, arXiv:1312.4889

\bibitem[Cole et al.(2005)]{Cole05}
Cole, S., et al., 2005, MNRAS, 362, 505

\bibitem[Eisenstein \& Hu(1998)]{EH98}
  Eisenstein, D.~J.; and Hu, W., ApJ, 496, 605 (1998)

\bibitem[Eisenstein, Seo, \& White(2007)]{Eisen07}
Eisenstein, D. J.; Seo, H.-J.; White, M. 2007, ApJ, 664, 660

\bibitem[Fixsen(2009)]{Fixsen09}
Fixsen, D.J. 2009, ApJ, 707, 916

\bibitem[Frieman, Turner, \& Huterer(2008)]{Frieman08}
Frieman, J., Turner, M., Huterer, D., ARAA, 46, 385 (2008)

\bibitem[Guzzo et al.(2008)]{Guzzo08}
Guzzo L. et al. 2008, Nature 451, 541

\bibitem[Hamilton(1992)]{Hamilton92}
Hamilton, A.~J.~S., 1992, APJL, 385, L5

\bibitem[Hartlap, Simon, \& Schneider(2007)]{Hartlap07}
Hartlap, J., Simon, P.,  \& Schneider, 2007, P., A\&A 464, 399

\bibitem[Hemantha, Wang, \& Chuang(2013)]{Hemantha13}
Hemantha, M. D. P.; Wang, Y.; Chuang, C.-H., arXiv:1310.6468

\bibitem[Kaiser(1987)]{Kaiser87}
Kaiser N., 1987, MNRAS 227, 1


\bibitem[Landy \& Szalay(1993)]{Landy93}
Landy, S.~D. and Szalay, A.~S. 1993, ApJ, 412, 64 

\bibitem[Laureijs et al.(2011)]{RB}
Laureijs R. et al. 2011, ``Euclid Definition Study Report'', arXiv:1110.3193


\bibitem[Lewis \& Bridle(2002)]{Lewis02}
Lewis, A. and Bridle, S., Phys.\ Rev.\  D {\bf 66}, 103511 (2002)


\bibitem[Li et al.(2011)]{Li11}
Li, M.; Li, X.-D.; Wang, S.; Wang, Y., 2011, Commun.Theor.Phys., 56, 525

\bibitem[Manera et al.(2013)]{Manera13}
Manera, M., et al. 2013, MNRAS, 428, 1036

\bibitem[Manera et al.(2015)]{Manera15}
Manera, M., et al. 2015, MNRAS, 447, 437

\bibitem[Mehta et al.(2012)]{Mehta12}
Mehta, K., et al., 2012, MNRAS, 427, 2168

\bibitem[Perlmutter et al.(1999)]{Perl99} 
Perlmutter S. et al. 1999, ApJ 517, 565

\bibitem[Press et al.(1992)]{Press92}
Press W.H., Teukolsky S,A., Vetterling W.T., Flannery B.P.,
1992, Numerical recipes in C. The art of scientific computing, 
Second edition, Cambridge University Press. 


\bibitem[Ratra \& Vogeley(2008)]{Ratra08}
Ratra, B., Vogeley, M.~S., 2008, Publ.Astron.Soc.Pac., 120, 235

\bibitem[Riess et al.(1998)]{Riess98}
Riess A. et al. 1998, AJ 116, 1009

\bibitem[Samushia et al.(2013)]{Samushia13}
Samushia, L., et al., 2013, arXiv:1312.4899

\bibitem[Sanchez, Baugh, \& Angulo(2008)]{Sanchez08} 	
Sanchez, Ariel G.; Baugh, C. M.; Angulo, R. 2008,
MNRAS, 390, 1470

\bibitem[Sanchez et al.(2013)]{Sanchez13}
Sanchez, A.G., 2013, arXiv:1312.4854

\bibitem[Scoccimarro(2004)]{Sco04}
Scoccimarro, R., 2004, Phys. Rev. D, 70, 083007

\bibitem[Seo \& Eisenstein(2003)]{Seo03}
Seo H., Eisenstein D. 2003, ApJ 598, 720

\bibitem[Song \& Percival(2009)]{Song09}
Song, Y.-S.; \& Percival, W.J. 2009,  JCAP, 0910, 004

\bibitem[Spergel et al.(2015)]{SDT15}
Spergel, D., et al., 2015, {\it Wide-Field InfrarRed Survey Telescope-Astrophysics Focused Telescope Assets WFIRST-AFTA 2015 Report},
eprint arXiv:1503.03757

\bibitem[Uzan(2010)]{Uzan10}
Uzan, J.-P. 2010, General Relativity and Gravitation, 42, 2219

\bibitem[Wang(2008)]{Wang08}
Wang Y. 2008, JCAP 0805, 021 

\bibitem[Wang(2010)]{Wang10}
Wang, Y., {\it Dark Energy}, Wiley-VCH (2010)

\bibitem[Wang \& Wang(2013)]{Wang13}
Wang, Y.; Wang, S. 2013, Phys. Rev. D 88, 043522

\bibitem[Wang, Chuang, \& Hirata(2013)]{WCH13}
Wang, Y.; Chuang, C.-H.; Hirata, C.M., 2013, MNRAS, 430, 2446

\bibitem[Wang(2014)]{Wang14}
Wang, Y., 2014, MNRAS, 443, 2950

\bibitem[Weinberg et al.(2013)]{Weinberg12}
Weinberg, D. H.; Mortonson, M. J.; Eisenstein, D.J.; Hirata, C.; Riess, A. G.; Rozo, E
2013, Physics Reports, 530, 87

\bibitem[Zhang, Pan, \& Zheng(2013)]{Zhang13}
Zhang, P.-J., Pan, J., and Zheng, Y., 2013, Phys. Rev. D, 87, 063526

\bibitem[Zheng et al.(2013)]{Zheng13}
Zheng, Y., Zhang, P.-J., Jing, Y-P., Lin, W.-P., \& Pan, J., 2013, Phys. Rev. D, 88, 103510

\end{thebibliography}
\end{document}